\pdfoutput=1
\documentclass[final,3p,times]{elsarticle}
\usepackage{pdfpages}
\usepackage{amssymb}
\usepackage{soul}
\usepackage{color}
\usepackage{xcolor}
\usepackage{wrapfig}
\usepackage{enumitem}
\usepackage{listings}
\usepackage{amsmath} 
\usepackage{color,colortbl}
\usepackage{algorithm}
\usepackage[noend]{algpseudocode}
\usepackage{todonotes}
\usepackage{braket}
\usepackage{caption}
\usepackage{graphicx, subfig}
\usepackage{dblfloatfix}
\usepackage{float}
\usepackage{comment}
\usepackage{mathtools}

\DeclarePairedDelimiter\floor{\lfloor}{\rfloor}
\usepackage{lipsum}
\usepackage{multicol}
\usepackage[T1]{fontenc}
\usepackage{epstopdf}

	\lstdefinestyle{customc}{
		breaklines=true,
		frame=none,
		xleftmargin=\parindent,
		language=Java,
		showstringspaces=false,
		basicstyle=\ttfamily\footnotesize,
		keywordstyle=\bfseries\color{green!40!black},
		commentstyle=\itshape\color{purple!40!black},
		identifierstyle=\color{blue},
		stringstyle=\color{orange},
	}
	\journal{Parallel Computing}

\begin{document}
\nocite{*}
\begin{frontmatter}
			\title{Client-side Straggler-Aware I/O Scheduler for \\ Object-based Parallel File Systems}
			\author{Neda Tavakoli}
			\author{Dong Dai}
			\author{Yong Chen}
			\address{neda.tavakoli@ttu.edu, dong.dai@ttu.edu, yong.chen@ttu.edu}
			\address{Computer Science Department, Texas Tech University}
%
\begin{abstract}
Object-based parallel file systems have emerged as promising storage solutions for high-performance computing (HPC) systems. Despite the fact that object storage provides a flexible interface, scheduling highly concurrent I/O requests that access a large number of objects still remains as a challenging problem, especially in the case when stragglers (storage servers that are significantly slower than others) exist in the system. An efficient I/O scheduler needs to avoid possible stragglers to achieve low latency and high throughput. In this paper, we introduce a log-assisted straggler-aware I/O scheduling to mitigate the impact of storage server stragglers. The contribution of this study is threefold. First, we introduce a client-side, log-assisted, straggler-aware I/O scheduler architecture to tackle the storage straggler issue in HPC systems. Second, we present three scheduling algorithms that can make efficient decision for scheduling I/Os while avoiding stragglers based on such an architecture. Third, we evaluate the proposed I/O scheduler using simulations, and the simulation results have confirmed the promise of the newly introduced straggler-aware I/O scheduler.
\end{abstract}
\end{frontmatter}
%
\section{Introduction}
\begin{figure}[h!]
	\begin{center}
		\includegraphics[width=3.5in]{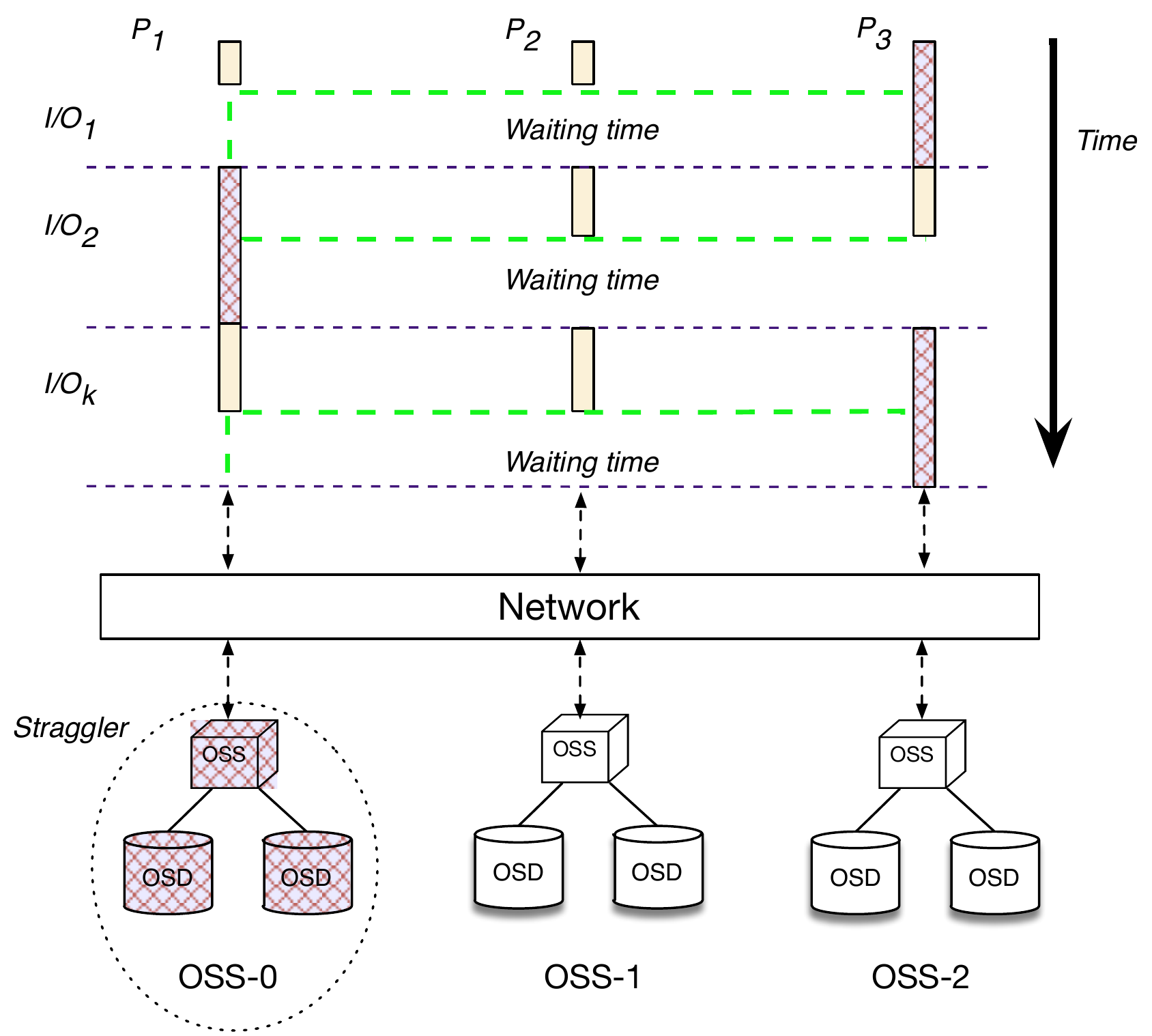}
		\caption{An illustration of how a straggler in object storage servers can affect the I/O performance in HPC.}
		\label{straggler_example}
	\end{center}
\end{figure}

The shift toward data-driven scientific discovery and innovation has made many high-performance computing (HPC) applications more highly data intensive than ever before. 
The I/O performance is considered an increasingly critical factor that determines the overall HPC system performance. In the meantime, object-based parallel file systems~\cite{mesnier2003object}, in which files are represented as a set of objects stored on object-based storage devices (OSDs)~\cite{liu2011towards,del1993improved}, managed by object storage servers (OSSs), have been increasingly deployed on large-scale high-end/high-performance computing systems due to their merits of improved scalability, manageability, and performance~\cite{dai2012sedna,factor2005object}, especially when highly concurrent I/O requests occur.
Developing a highly efficient I/O scheduler in the object-based storage systems is arguably critical and a well-acknowledged challenge. A considerable amount of work has been done in this space~\cite{jain1997heuristics,chen2001performance,durand2003parallel,liu2011towards,rosti1998impact,wiseman2003paired,rosti2002models}. Among them, the \textit{straggler} problem~\cite{wang2004obfs,thakur1999data, xie2012characterizing,ousterhout2013case,ananthanarayanan2010reining}, which occurs when some of OSSs take a much longer time in responding to I/O requests than other servers, has drawn lots of attention recently. 
The occurrence of stragglers has significant effects on I/O performance of object storage systems. Since in HPC applications, clients normally need to synchronize after each I/O phase~\cite{del1993improved,liu2015hierarchical}, the overall I/O performance will be determined by the longest one, which in turn is determined by the {\em slowest} object storage server. 
In general, the slow storage servers (i.e., stragglers) can be divided into two categories: long-term stragglers and short-term stragglers. Long-term stragglers can be slow for hours, days and even in the worst-case forever. The main reasons of occurring this type of straggler can be outdated hardware, hardware failures and even software bugs. On the other hand, short-term stragglers last only for minutes or less. The main reasons of this type of straggler are interference or resource contention between applications.
Figure~\ref{straggler_example} illustrates an example of how a straggler in object storage servers can affect the I/O performance. Assuming application processes issue three I/O requests, and each line on the top of the figure represents part of each I/O to access OSSs. The bottom part of the figure illustrates three OSSs in this example, assuming that OSS-0, in shaded pattern, is a straggler. If any part of an I/O hits the straggler server OSS-0, the entire I/O suffers from the straggler problem because the I/O cannot be completed until the slowest server OSS-0 finish servicing its part of I/O access. The existence of storage server stragglers can significantly reduce the productivity of the HPC system.
Even though an asynchronous I/O can decouple the computation and I/O and is helpful in some cases, but it is not always possible. For instance, if the computation depends on the I/O reads, or the simulation output is so large that these data have to be flushed to storage systems before the computation can carry on, the application processes have to be blocked until the I/O is completed. The storage server straggler problem can be catastrophic in projected extreme-scale systems, as the large-scale storage system significantly increases the possibility of the existence of a straggler in storage servers.

In order to mitigate the impact of storage server stragglers, a straggler-aware I/O scheduling mechanism is imperatively desired for increasingly popular object-based parallel file systems. 
Numerous recent studies have focused on this problem, including our prior research of a two-choice randomized I/O scheduling strategy~\cite{dai2014two}. However, these strategies suffer from expensive 
probing messages and communication overhead, as they need to probe storage servers' status and outstanding workloads to detect the existence of stragglers and adopt a randomized scheduling strategy to avoid the straggler.
In this research study, to overcome the limitation of probing message based straggler-aware I/O scheduling, we introduce a client-side, {\em log-assisted straggler-aware I/O scheduling}. The fundamental idea is that, the I/O scheduler at the client side maintains a log of I/O requests, server's status, and past scheduling decisions, so that the I/O scheduler client can make an optimized scheduling decision when stragglers exist, without incurring expensive probing messages and communication cost. Based on this new idea, we also introduce three different scheduling policies that leverage this log information to make the scheduling decision. To validate the newly introduced log-assisted straggler-aware I/O scheduling strategy and to verify the effectiveness of different scheduling policies, we have conducted simulation evaluation, and these simulation results confirm that the newly proposed scheduler is effective in improving I/O performance with avoiding storage server stragglers and significantly reducing probing messages and communication cost.
The contribution of this research study is three-fold: 
\begin{itemize}
	\item Introduce a client-side, log-assisted straggler-aware I/O scheduling for object-based parallel file systems with the log design and the mechanism of maintaining the logs at the scheduler client;
	\item  Introduce three scheduling algorithms based on the log-assisted straggler-aware scheduling mechanism that can be used in parallel file systems;
	\item Conduct simulation evaluations to validate the concept and analyze the effectiveness of a log-assisted straggler-aware scheduler and two different scheduling algorithms.
\end{itemize}
The rest of this paper is organized as follows. Section~\ref{s2} gives the background and motivation of this study. Section~\ref{s3} gives a detailed description of the proposed client-side log-assisted I/O scheduler and proposed scheduling algorithms. 
Section~\ref{s4} reports the experimental results. Section~\ref{s5} discusses relevant work and compares with this study. We conclude this research and describe possible future work in  Section~\ref{s6}.
%
\section{Background \& Motivation}
\label{s2}
\subsection{HPC I/O Stack}
\begin{figure}[H]
	\begin{center}
		\includegraphics[width=1.5in]{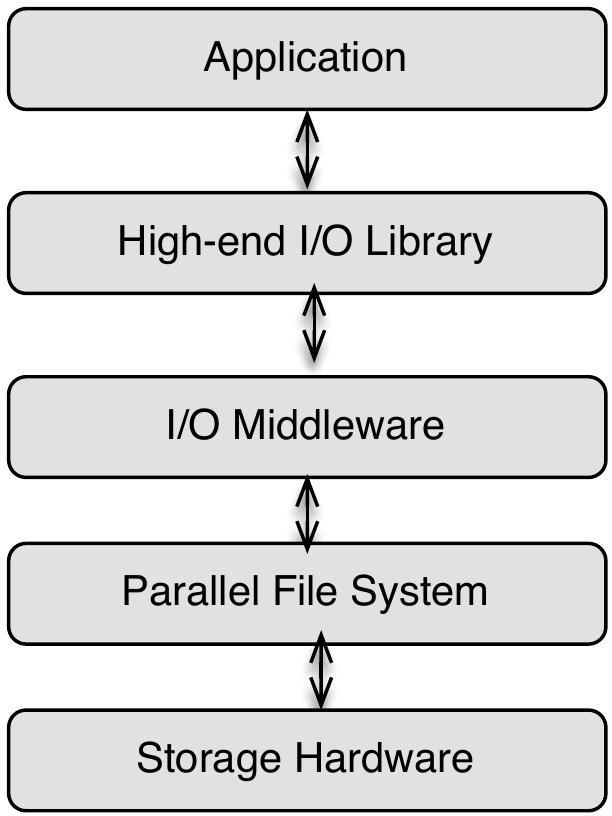}
		\caption{HPC I/O software stack.}
		\label{f1}
	\end{center}
\end{figure}

\begin{figure}[H]
	\begin{center}
		\includegraphics[scale= 0.4,width=0.4\textwidth]{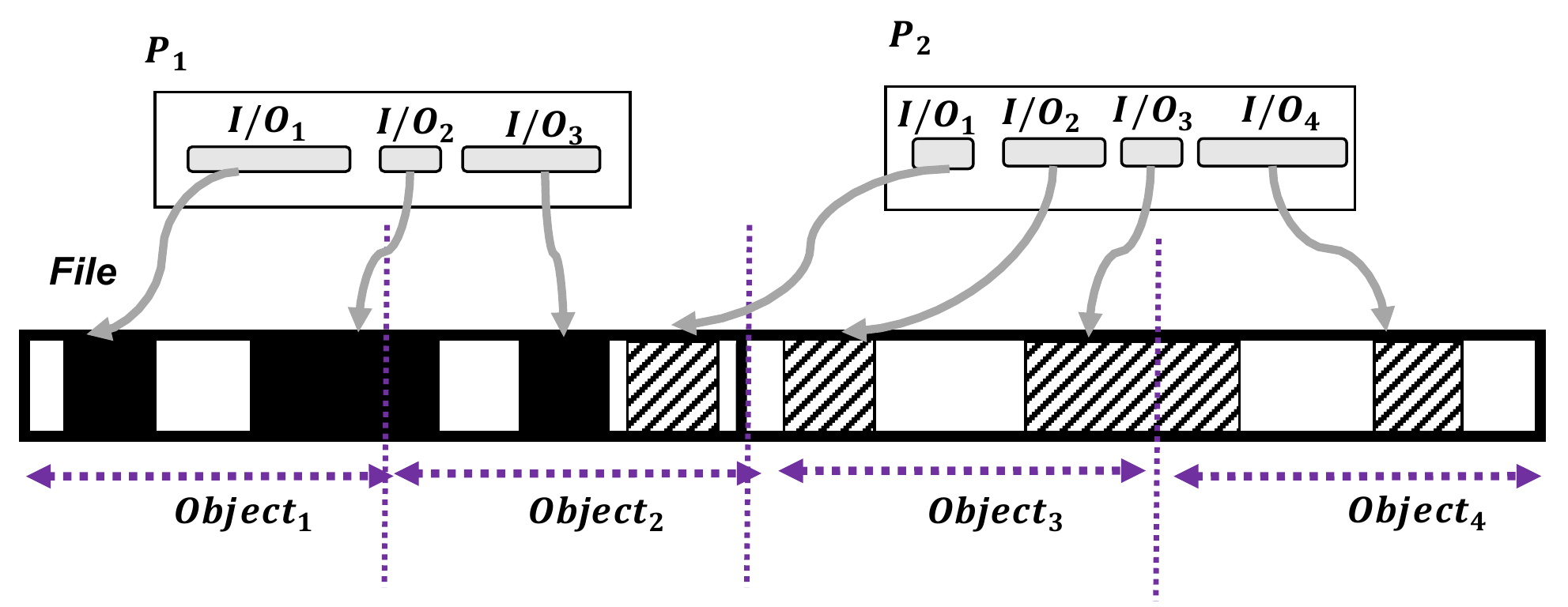}
		\caption{ Example of object file.}
		\label{fig2}
	\end{center}
\end{figure}

In Figure~\ref{f1}, we illustrate a typical HPC I/O software stack. The hierarchical view starts from the \textit{application} level, which issues the I/O requests through underlying libraries. There are various \textit{I/O libraries} that can be used for applications to easily and efficiently describe their data requests, for examples HDF5 \cite{borthakur2008hdfs}, parallel NetCDF \cite{li2003parallel} and ADIOS \cite{liu2014hello}. In most cases, these libraries will translate users' requests into function calls to the \textit{I/O middleware} such as MPI-IO \cite{thakur1999implementing}. The I/O middleware normally provides performance optimizations like collective I/O  and data sieving \cite{thakur1999data}. All the requests are finally translated into POSIX function calls to underlying \textit{parallel file system} such as GPFS \cite{schmuck2002gpfs}, Luster \cite{braam2004lustre}, PVFS, etc.
Together with the rapid adoption of object-based storage systems (storage hardware, e.g. object storage devices, OSDs), most parallel file systems in the I/O software stack also become object-based. 
Along with those I/O software stack layers, the scheduler can run on several places. It can run in high-end I/O libraries~\cite{liu2014hello}, in I/O middleware~\cite{dorier2014calciom}, or even in the client-side library of parallel file systems~\cite{song2011server}. In this paper, we focus on the last case, i.e., the client-side library of parallel file systems. 
In object-based parallel file systems, it is actually the client library to divide a file into multiple objects and issue requests on those objects. Figure~\ref{fig2} illustrates an example of how an object-based parallel file system schedules I/O requests. Normally, there will be multiple applications running with multiple processes to issue object access requests. In this figure, we only show two processes as an example. Here, $p1$ and $p2$ running on different compute nodes. Each of them issues multiple I/O requests on different objects. For example, $I/O_1$ of $ p_1$ accesses $ object_1$ and $I/O_4$ of $ p_2$ accessing $ object_4$.
I/O requests may hit the boundary of objects, so it will be split into several requests. For example, $I/O_2$ of $ p_1$ accessing both $ object_1$ and $ object_2$. Each object will be scheduled separately. In the following discussion, we no longer explicitly differentiate this scenario for simplicity. 
After scheduling, each I/O request will be assigned to an appropriate OSD, which is the decision made by the I/O scheduler of object-based parallel file systems. This OSD will provide service for accessing the object.
%
\subsection{Dynamic I/O Scheduler}
In previous work \cite{dai2014two}, we have proposed a dynamic client-side I/O scheduler for object storage systems in HPC environment. The core idea of such a scheduler is to dynamically schedule I/O requests to the best appropriate server-based on their real-time performance instead of always scheduling certain object to a fixed object server.  Figure 4 shows how the proposed I/O scheduler works for read and write requests as an example.
\begin{figure}[H]
	\begin{center}
		\includegraphics[scale= 0.4,width=0.5\textwidth]{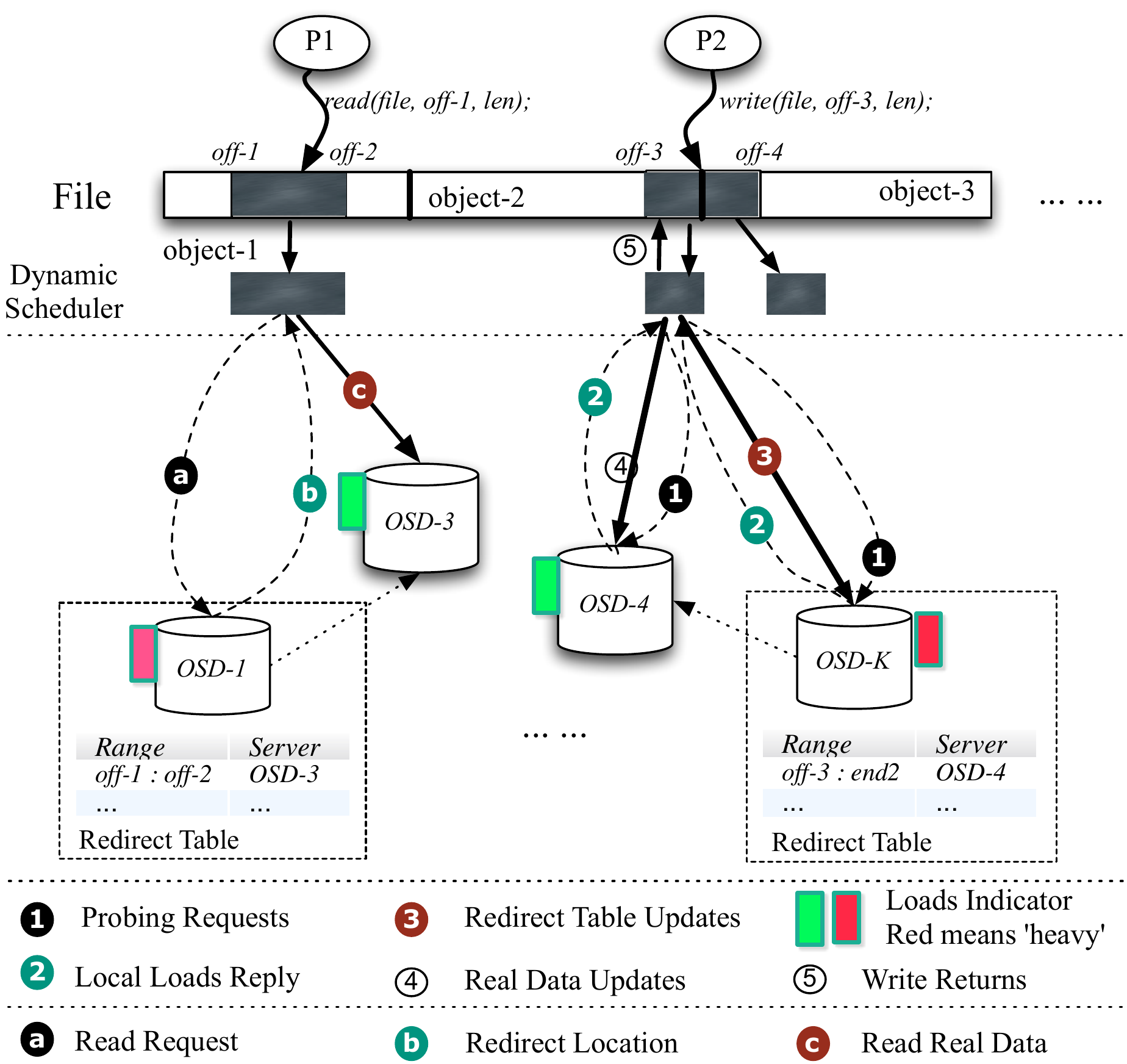}
		\caption{ The procedure of dynamic I/O schedulers\cite{dai2014two}.}
	\end{center}
\end{figure}
Each time when a new write request is issued on a certain object, the scheduler will first randomly probe two servers (one of them will be original server) and choose the best performing one as the candidate for serving the real I/O requests. This approach can effectively avoid possible stragglers since it compares two object servers each time, hence has a high possibility to avoid stragglers. But, as the figure shows, it may also create overhead on the future read operations, which need to first find the location and then issue the read.
Although there are many advantages in avoiding stragglers, dynamic scheduling does introduce other problems. First, it will introduce extra overhead due to the probe phase before I/O scheduling each time. We have proposed a collaborative probe strategy to collaboratively collect performance of storage servers to reduce such overhead. 
Second, if multiple clients try to probe at the same time, they will get the same best performing server and lead to a possible new bottleneck (i.e., straggler) because of the scheduling action. We have introduced a pre-assigned strategy to tackle this. By pre-assigning, each server will predict whether it will be chosen or not based on historical data and current load. If a server predicts itself to be highly possibly chosen, later probes from other clients will receive updated loads.

Although the two-choice randomized scheduler tends to achieve more balanced workload distribution and better performance when stragglers exist, it still can be largely improved. One significant area is to improve the accuracy and efficiency of client-side probing. It is obvious that client-side schedulers actually are able to learn OSD statuses by recording the historical information about previous scheduling decisions and their performance. In this research, we follow the dynamic I/O scheduling approach, but introduce a new client-side logging mechanism to avoid the probing.  
%
\section{Architecture and Design}
\label{s3}
\subsection{System Architecture}
\begin{figure}[H]
	\begin{center}
		\includegraphics[scale= 0.4,width=0.5\textwidth]{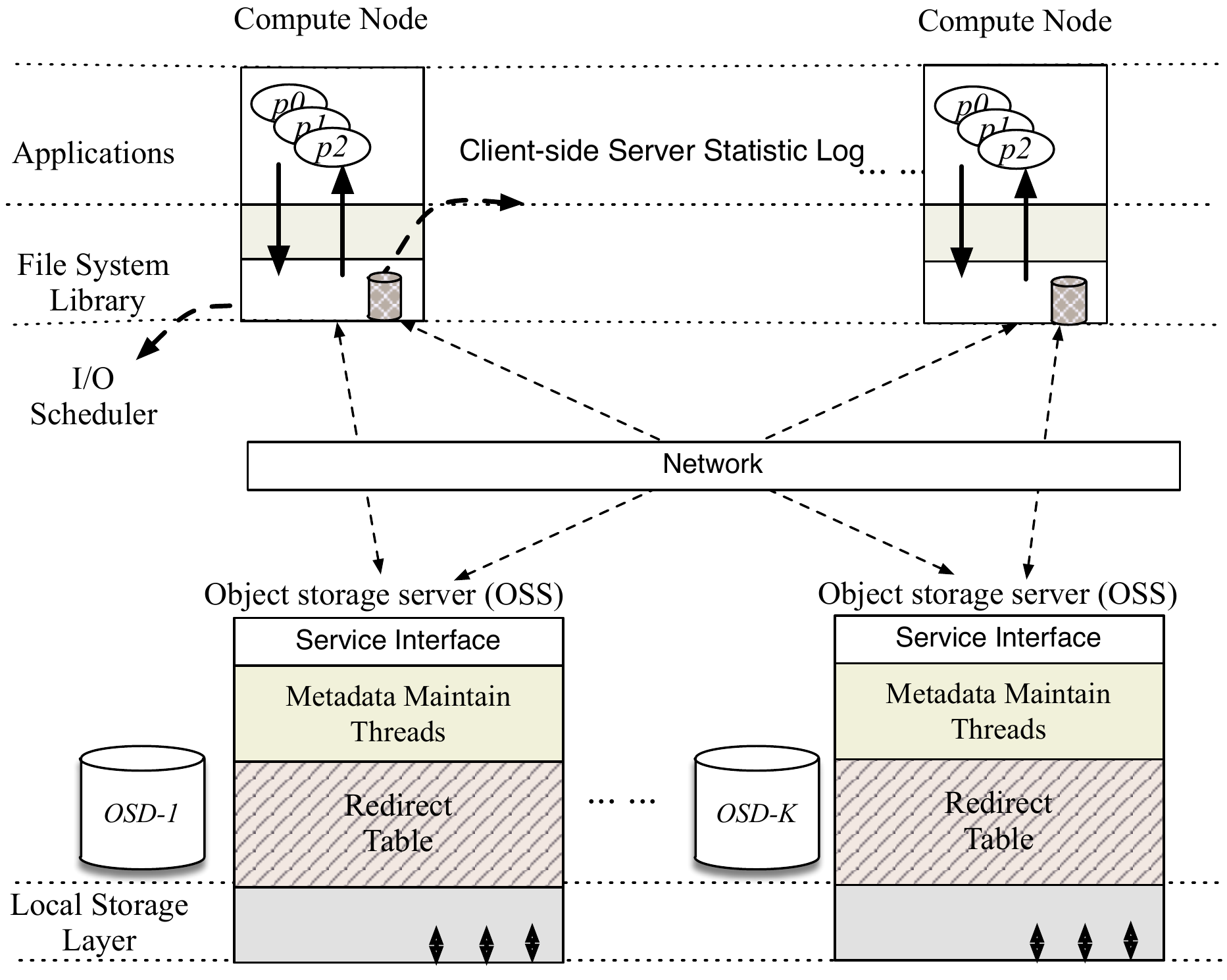}
		\caption{Overall system architecture of a log-assisted straggler-aware I/O scheduler.}
		\label{arch}
	\end{center}
\end{figure}

\begin{figure}[H]
	\begin{center}
		\includegraphics[width=2.8in,height=1.2in]{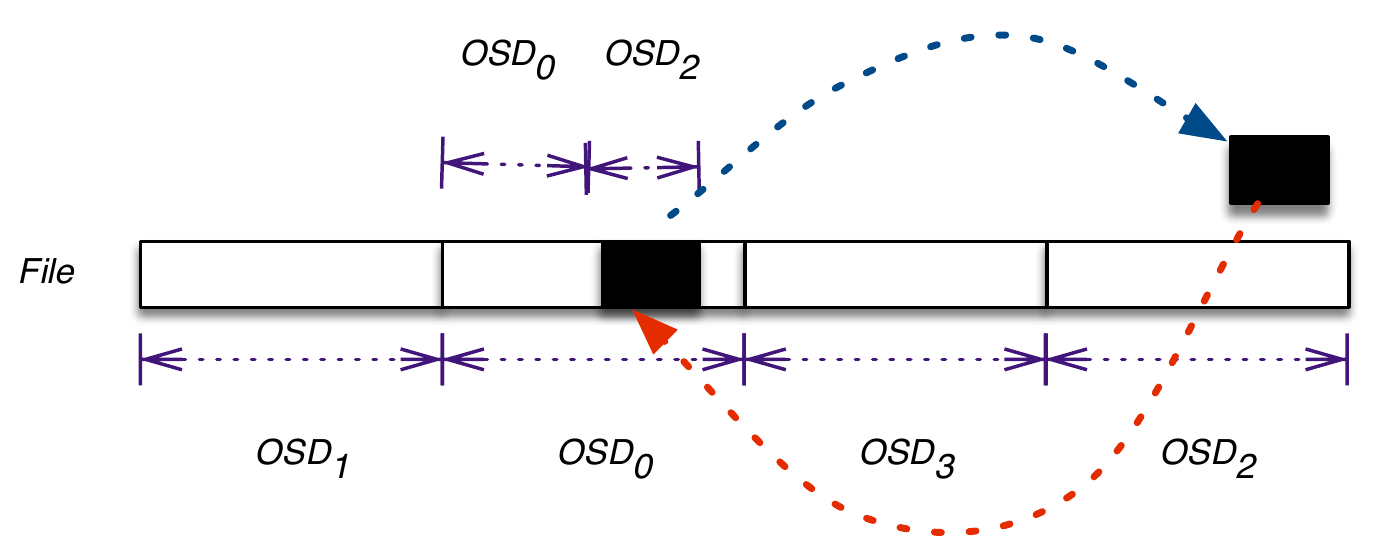}
		\caption{Redirect table for straggler-aware I/O scheduler.}
		\label{redirect}
	\end{center}
\end{figure}

The overall architecture of the proposed log-assisted straggler-aware I/O scheduler is illustrated in Fig.~\ref{arch}.  
The proposed scheduler runs on the client-side (compute nodes), as an I/O scheduler serving each I/O request. A key component of the proposed scheduler is a \textit{client-side server statistic log}, which will be described in detail in the next subsection.
In the object storage server side, we reuse the components designed and implemented in our previous work~\cite{dai2014two}. Specifically, we have a \textit{redirect table} and \textit{metadata maintainer thread} running on the object storage servers. Among them, the redirect table is used to remember the default location of data objects. This is necessary since we will dynamically place I/O requests to different object storage servers based on the scheduling decision. Obtaining a distributed redirect table in each object storage server is more scalable compared with storing every redirection into metadata service of the parallel file systems. Another component, the metadata maintainer thread runs in the background to move the redirected objects back to their default location when the file systems are idle. This helps keep the data location consistent with metadata in parallel file systems and hence improves the performance of further reads. 
Fig.~\ref{redirect}  further illustrates an example of how the redirect table and metadata maintainer thread work. It shows a data fragment (in black), which should be placed on $OSD_0$ but is scheduled to be written to $OSD_2$ as $OSD_0$ is found to be a straggler. 
After the scheduler arranges the data fragment to be written to $OSD_2$, a new entry will be created in the redirect table of $OSD_0$ indicating that the current location of the segment is on $OSD_2$. The dashed red line indicates the metadata maintainer thread periodically runs to retrieve this segment back and deletes the entry in the redirect table. More details can be found in our previous work~\cite{dai2014two}.
%
\subsection{Scheduling Model}
In the proposed architecture, for each application run with multiple processes, the I/O requests from each process are combined and scheduled to access object storage servers together in a fashion similar to the collective I/O. These I/O requests are not necessarily from the same process and can be from a group of processes, similar as in collective I/O. 
%
Specifically, the concurrent I/O requests issued in one client are queued temporarily and served in the group. We define these multiple I/O requests served together as a \textit{time series}, similar as in~\cite{song2011server}. 
%
\begin{figure}[H]
	\begin{center}
		\includegraphics[scale= .5]{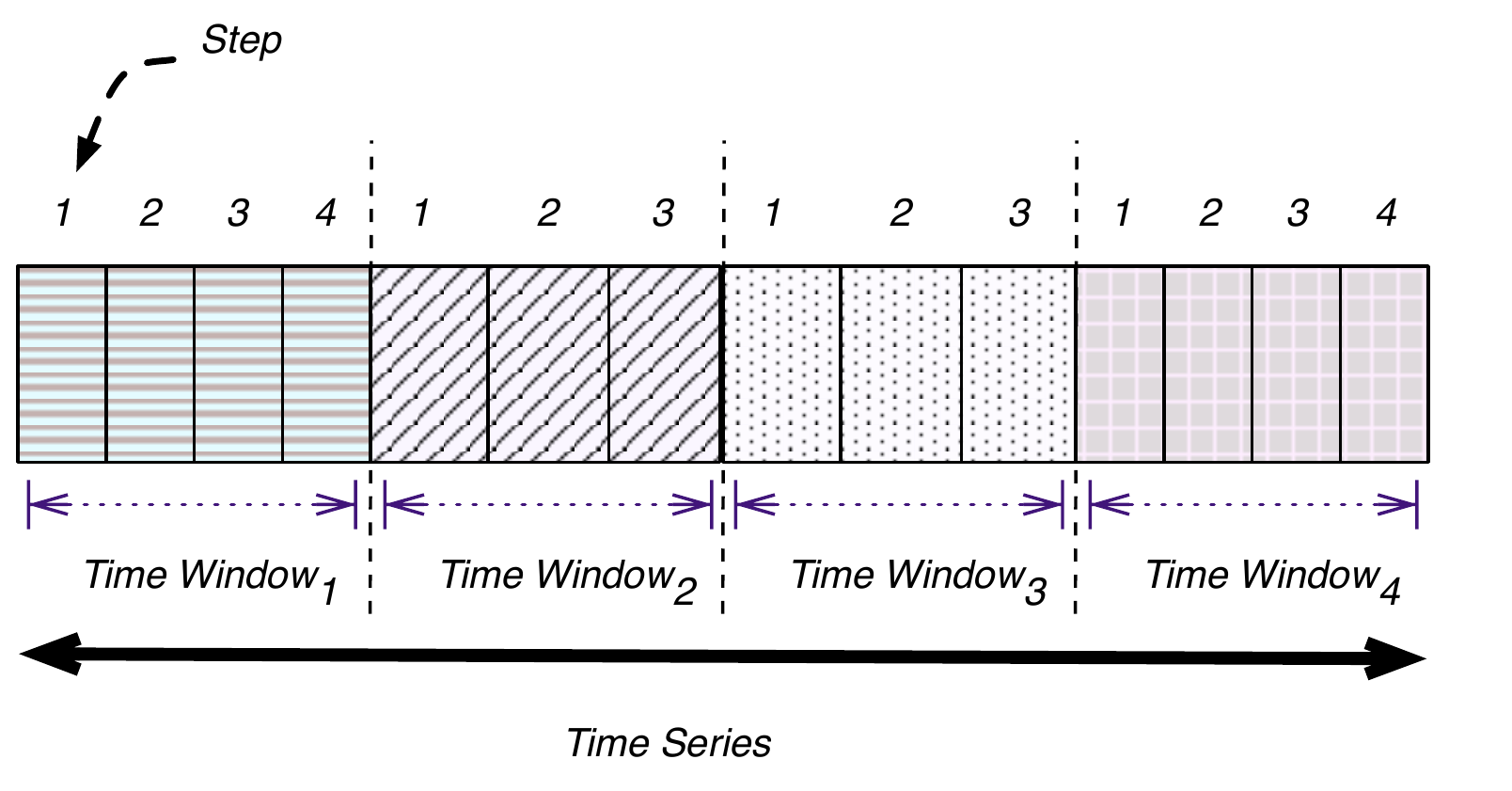}
		\caption{Time series, time window, and step number in scheduling.}
		\label{timewindow}
	\end{center}
\end{figure}
We show the concept of time series in Fig.~\ref{timewindow}. Time series is divided into consecutive segments called \textit{time window}, which contains a fixed time interval. Each time, the proposed I/O scheduler will serve all buffered requests before moving to the next time window.
Inside one time window, a number of I/O requests can be queued. We divide queued I/O requests in each time window into multiple steps. Each step will contain the I/O requests on the same object to reduce unnecessary network communications. 
%
\subsection{Client-side Server Statistic Log}
A client-side server statistic log is introduced as a core component of a new log-assisted straggler-aware I/O scheduler. As shown in Fig.~\ref{arch}, it works inside the I/O scheduler on the client-side.

\subsubsection{Data Structure of Client-Side Server Statistic Log}
\begin{figure}[H]
	\begin{center}
		\includegraphics[scale=0.7,width=0.4\textwidth]{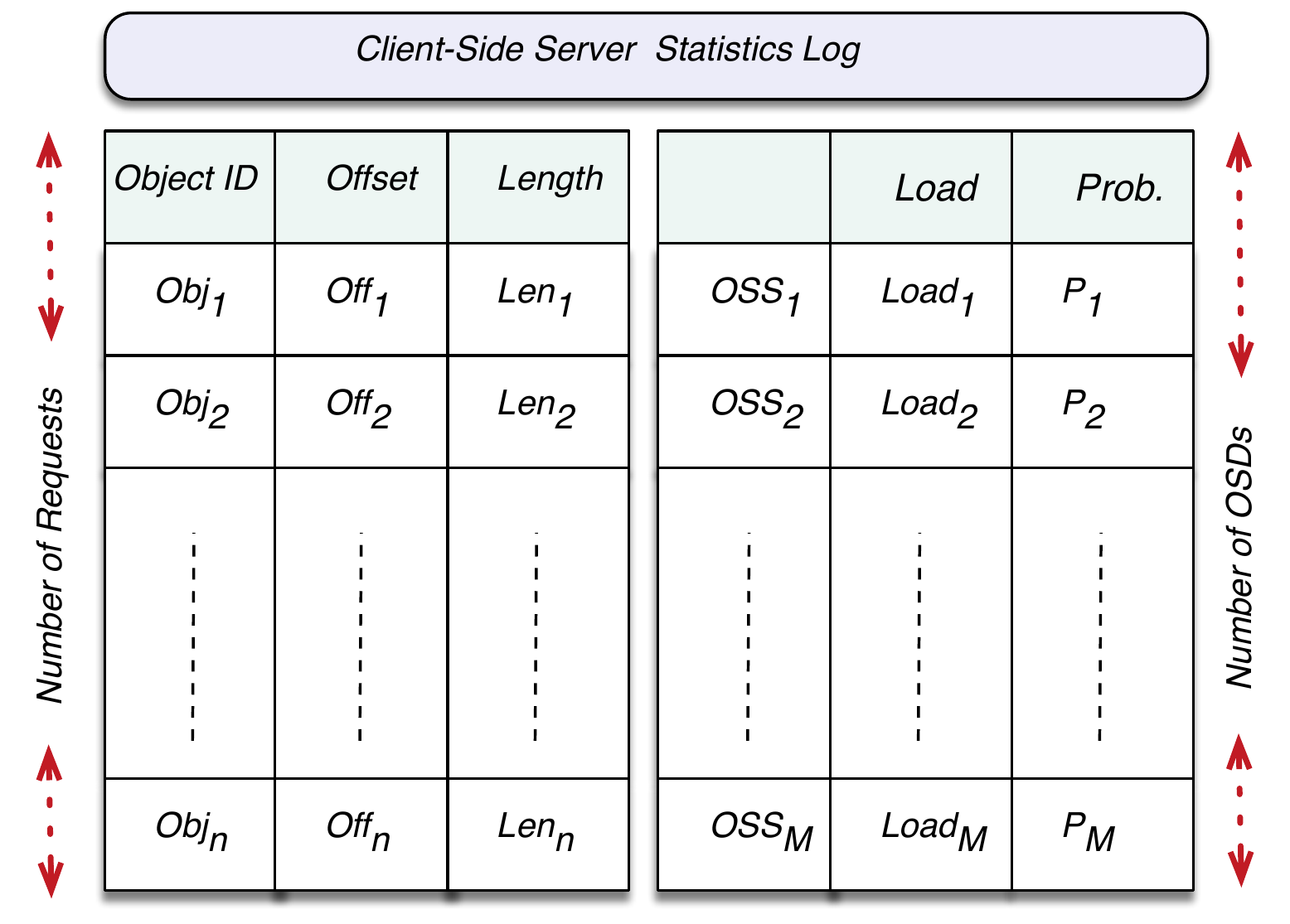}
		\caption{ Client-side server statistic log.}
		\label{log}
	\end{center}
\end{figure}
We show the detailed structure of the client-side server statistic log in Fig.~\ref{log}. It consists of two tables: an \textit{I/O request table}, shown as the left part of Fig.~\ref{log}, and a \textit{server statistic table}, shown in the right part.
In the I/O request table, each request forms a row, which consists of three fields: 1) \textit{Object ID}, which is a unique identifier (a very large index number) of each object; 2) \textit{Offset}, which denotes the starting offset of the I/O request from the beginning of the object; and 3) \textit{Length}, which is the size of the I/O request (in bytes).
On the other hand, the server statistic table consists of statistics of each object storage server. It is generated based on scheduling decisions from previous time windows. The \textit{Load} field is an expectation of the amount of I/O requests (in terms of bytes) in each object storage server. 
Different from our previous work~\cite{dai2014two}, in this research, we do not probe the servers to retrieve their realtime loads; instead, we use this server statistic table to update and record current load of each storage server based on previous scheduling decisions. We use such load to calculate the probability of selecting a corresponding storage server for a given I/O request, which is maintained and updated in the \textit{Prob} field. In the next subsection, we introduce how to update the server statistic table and also calculate the probability.
%
\subsubsection{Maintaining Servers Statistic Table}
First, after serving each I/O request in a single step, the server loads in the server statistic table will be updated based on the following formula:
\begin{equation}
l = l' + Len
\label{e1}
\end{equation}
where $l$ is the expected server load at the end of each step, $l'$ represents the server load from the previous step, and $Len$ is the length of scheduled I/O requests.
At the beginning of the next step, we will calculate the probability of selecting server $i$ and use such probability to choose the best storage server to schedule the I/O request. 
The probability is calculated according to the following formula:
\begin{equation}
p_i=p'_{i} * e^{-l_i}
\label{e2}
\end{equation}
where $p'_{i}$ indicates the current probability of selecting server $i$, and $l$ equals to the updated load of server $i$. 
The initial $p'_i$ is calculated based on the default round-robin scheduling strategy: given there are $M$ object storage servers, each server is equally assigned a $p'_i=\frac{1}{M}$.
Through calculating such a probability for each object storage server, we are able to choose one as target (the detailed selection algorithms are introduced in the next subsection). After making a scheduling decision, we will need to update the server statistic table based on Equation~\ref{e1}. 
In addition to updating the load of the selected server, we also need to update the probability of choosing another server to prepare for the next scheduling. In fact, the summation of probabilities of choosing all storage servers should equal to $1$. This also requires us to update their probabilities too. The following equation shows how this is done:
\begin{equation}
p_j = p_{j}^{'}+ \frac {(p'_i - p'_i * e^{-l_i})}{M-1},   j\neq i
\end{equation}
where $M$ indicates the total number of storage servers, $i$ is the server that is chosen to serve the current I/O request, and $j$ refers to all other object storage servers. This equation indicates that we evenly re-distribute the decreased probability of server $i$ to all other object storage servers ($j$). We use this strategy to maintain the probability.
Note that, in Equation~\ref{e2}, we use exponential distribution to calculate the probability of a server being chosen according to their current loads. We use the exponential distribution \cite{marshall1967multivariate} to calculate it because this distribution describes the behavior of balanced scheduling. Specifically, it indicates that servers with high loads (which could be very high due to imbalanced requests) should be considered with lower probability to serve incoming I/O requests to avoid the straggler. On the other hand, the server with lighter loads (which could approach to $0$) should be considered with a much higher probability of being chosen. If a server does not contain any load (i.e., $0$), it should be selected with the highest priority. Other distributions can possibly describe similar behaviors; however, since that is not our focus in this research, we leave the investigation of distributions as a future work.
%
\subsection{Scheduling Algorithms}
Based on the information stored in the server statistic log, we propose three new scheduling algorithms to leverage this statistic log to develop an efficient straggler-aware I/O scheduler.
Before introducing the proposed scheduling algorithms, we first introduce the base-line algorithm, a round-robin (RR) algorithm. This strategy is widely used in modern parallel file systems like PVFS, GPFS~\cite{schmuck2002gpfs}, and Lustre~\cite{halbwachs1991synchronous}. 
For an I/O request on object $i$, RR schedules it to the storage server $i \mod M$, where $M$ indicates the total number of object storage servers to be fair among multiple I/O requests and servers, as well as to avoid starvation.
It is not trivial to see that the RR strategy does not work well with stragglers as it does not consider current performance of storage servers. Hence, it may still choose the overloaded servers even there are lightly-loaded servers that can be chosen.

We introduce three straggler-aware scheduling algorithms based on the proposed client-side server statistic log. The first one, called \textit{Max Length - Min Load (MLML)} scheduling algorithm, directly uses logged server statistics to direct scheduling; i.e., it will select the server with lighter workloads with a higher probability.
The second scheduling algorithm, called \textit{Two Random from Top Half (TRH)}, takes advantage of the random strategy in addition to server statistic log.
The third one, called \textit{$n$-Level Two Random (nLTR)} scheduling algorithm,  which not only takes advantage of the random strategy but also classifies the storage servers (i.e, by using server static log) to different sections, which helps to spread the requests to be more balanced.

Our evaluations confirm that these three scheduling algorithms can reduce the impact of existing stragglers and avoid generating new stragglers. More details of their comparisons can be found in the evaluation section. We discuss each of these three algorithms in detail below.
%
\subsubsection{Max Length - Min Load (MLML) scheduling algorithm}
This scheduling strategy uses the statistic server log to sort the storage servers based on their probabilities from the highest to the lowest. At the same time, the I/O requests are sorted from the maximum length to the minimum one. These two sorted lists are processed,
from the top to the bottom of lists then starting again from the top of the lists in a circular manner. 
As a result, the maximum length I/O request will be scheduled to the server with lighter load and the minimum length I/O request will be distributed to the server with heavier load.

\begin{figure}[H]
	\begin{center}
		{\includegraphics[scale= 0.8,width=0.45\textwidth]{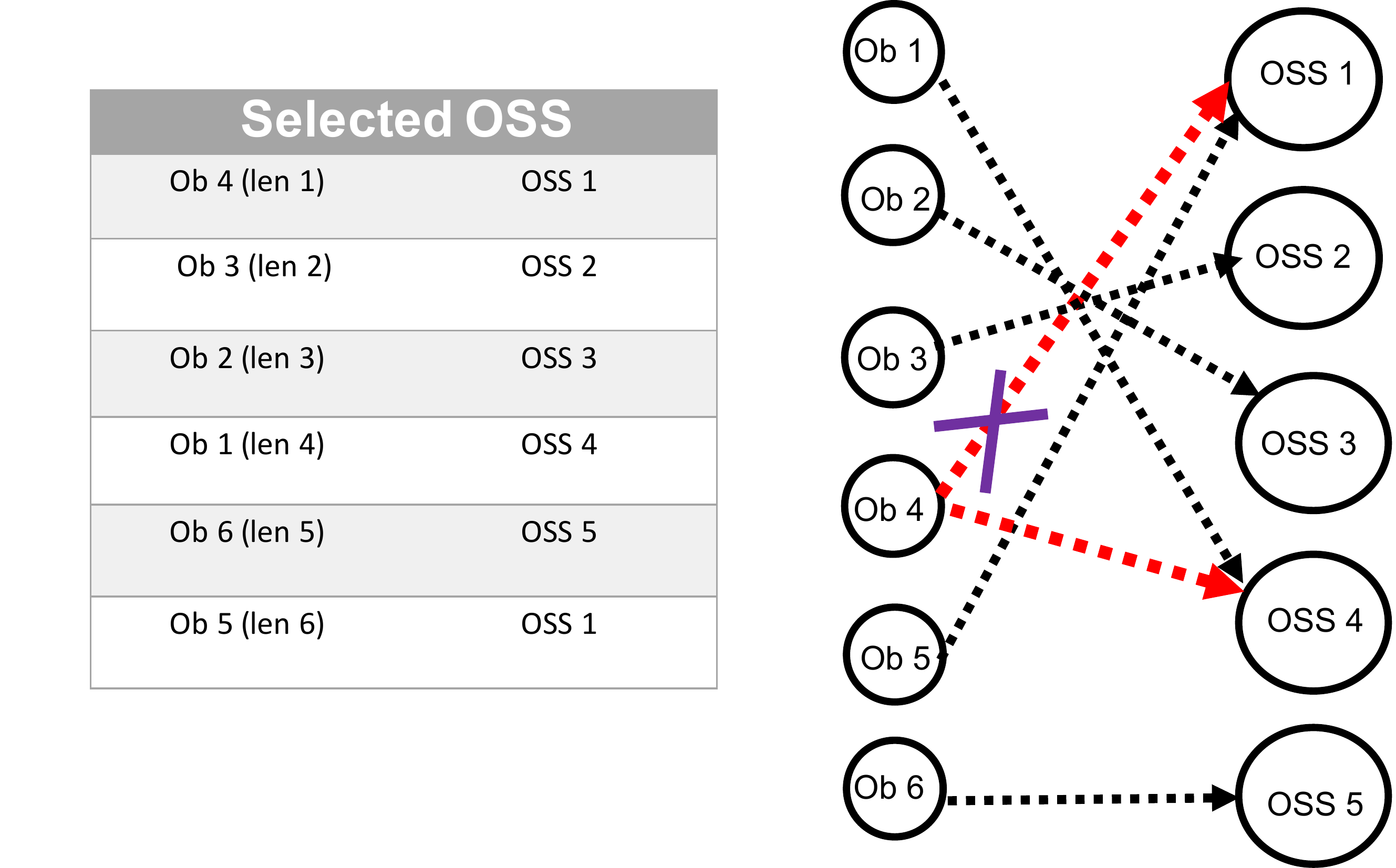}}
		\caption{ Example of MLML algorithm.}
		\label{MLML}
	\end{center}
\end{figure}

\begin{algorithm}
	\caption{Max Length-Min Load algorithm (MLML)}\label{1}
	\begin{multicols}{2}
		\begin{algorithmic}[1]
			\Procedure{MLML{(servers,requests,threshold)}}{}
			\State  sortedServers  = sort(servers);
			\State  sotRequests = sort(requests);
			\State index=0;
			\While{need\_schedule()}
			\State default\_oss = requests[index] $\mod$ M;
			\State target\_oss=sotRequests[index] $\mod$ M;
			\State benefit = load(default\_oss) - load(target\_oss);
			\If {benefit $\leq$ threshold}
			\State $\Return$ target\_oss;
			\Else
			\State $\Return$ default\_oss;
			\EndIf
			\State index++;
			\EndWhile
			\EndProcedure
		\end{algorithmic}
	\end{multicols}
\end{algorithm}

We describe the pseudocode of the MLML scheduling strategy in Algorithm~\ref{1}. In this algorithm, two sorted lists \textit{sortedServers} and \textit{sortedRequests} store a sorted list of servers and I/O requests, respectively.
The default$\_$oss describes the default location of requested objects. As we have described, such a server is selected through the default RR strategy. The target$\_$oss, on the other hand, represents the storage server calculated based on this MLML scheduling algorithm. 
If the target$\_$oss and the default$\_$oss are not the same server, we then need to consider whether it is worth of scheduling this request to the target$\_$oss as this might increase read overhead due to the redirection. We introduce a user-defined threshold to indicate how much overhead is acceptable.
Specifically, we compare loads of those two servers and if the benefit of choosing the target$\_$oss over the default$\_$oss is larger than the threshold, the selection is acceptable; otherwise the default server will still be selected. 
The rationale behind this algorithm is that a server with lighter load has a lower chance of being a straggler. By scheduling larger requests on the lighter server, we can reduce or minimize the impact of existing stragglers while the chance of generating new stragglers can be significantly decreased. 
Fig.~\ref{MLML} illustrates an example of how this MLML algorithm works.
In this example, there are six objects and five object storage servers. As shown in the figure, I/O requests are sorted based on their length, and OSSs are sorted based on their probability of being stragglers from the highest to the lowest. Based on the MLML strategy, these I/O requests are assigned to OSSs based on the sorted list. In this example, all selected OSSs by the scheduling algorithm provide benefits greater than the threshold, except for the ob 4, which will be scheduled to the default server (i.e. distributed to OSS4 instead of OSS1). Note that the redirect table is used to keep track these writes and maintains metadata consistency.
%
\subsubsection{Two Random from Top Half (TRH) algorithm}

\begin{figure}[H]
	\begin{center}
		{\includegraphics[scale= 0.8,width=0.45\textwidth]{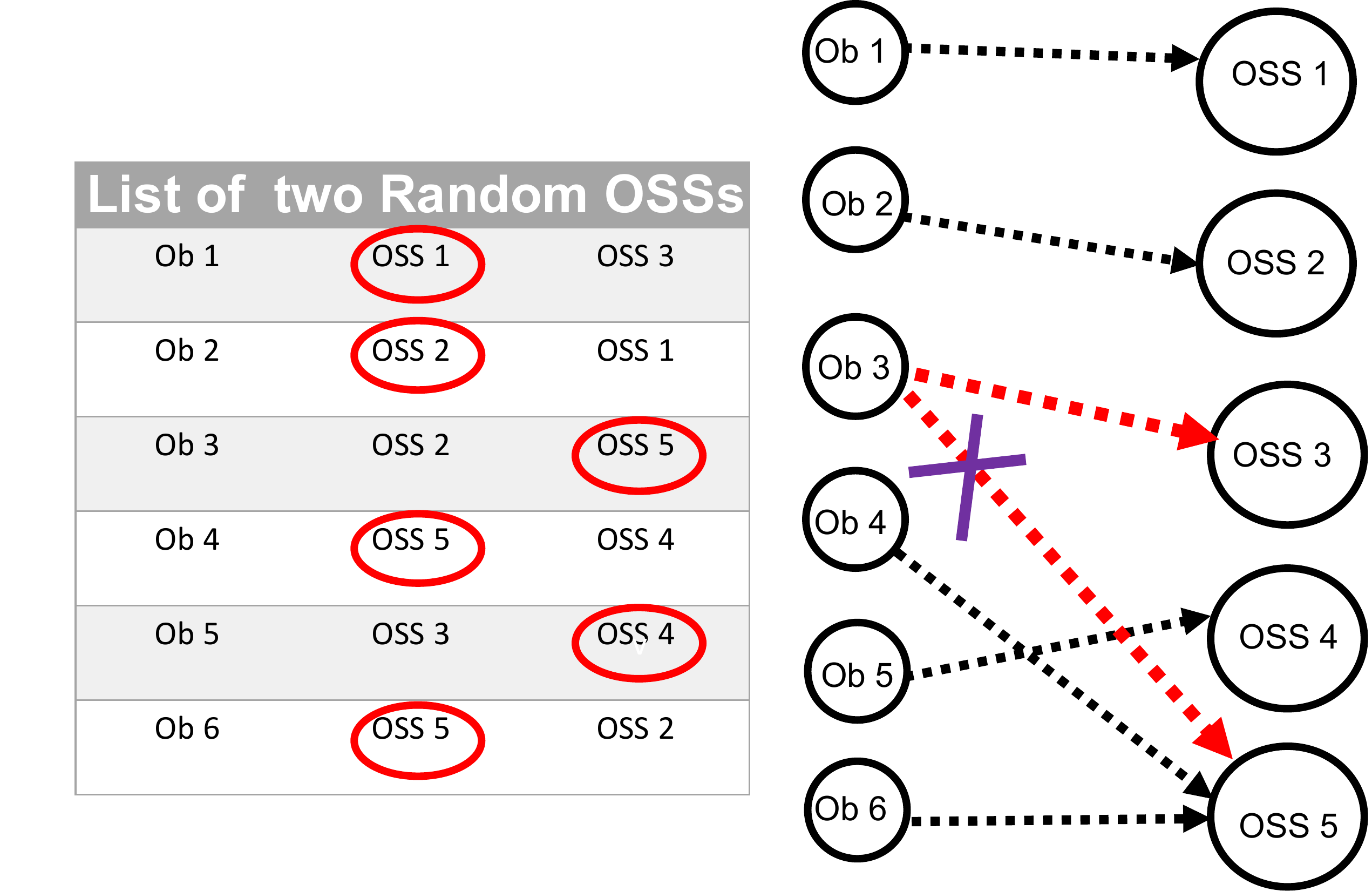}}
		\caption{ Example of TRH algorithm.} \label{TROME}
	\end{center}
\end{figure}

\begin{algorithm}
	\caption{Two Random from top Half (TRH) Algorithm}\label{2}
	\begin{multicols}{2}
		\begin{algorithmic}[1]
			\Procedure{TRH{(servers,requests,threshold)}}{}
			\State  sortedServers   = sort(servers);
			\State index=0;    
			\While{need\_schedule()}
			\State  default\_oss = requests[index] $\mod$ M;
			\State  ross\_1=random\_oss();
			\State  ross\_2=random\_oss();
			\State target\_oss =find\_min\_load(ross\_1, ross\_2);
			\State benefit = load(default\_oss) - load(target\_oss);
			\If{benefit $\leq$ threshold}
			\State{$\Return$ target\_oss;}
			\Else
			\State{$\Return$ default\_oss;}
			\EndIf
			\State index++;
			\EndWhile
			\EndProcedure
		\end{algorithmic}
	\end{multicols}
\end{algorithm}
This algorithm considers randomized choices based on the server statistic log to further mitigate the impact of stragglers. The rationale of using randomized choices in this algorithm is that it provides more possible options for selecting servers and helps to spread the requests to be more balanced. In this scheduling algorithm, the sorted list of object storage servers is also used, same as in the MLML algorithm.
Specifically, for a given I/O request, two object storage servers will be randomly chosen from the top half of all object storage servers. The top half indicates $\frac{M}{2}$ servers that have lighter loads. 
We also use the same threshold as in the previous algorithm to decide whether choosing the calculated target$\_$oss or the default$\_$oss. Algorithm~\ref{2} shows the pseudocode of the TRH algorithm. 

Fig.~\ref{TROME} further illustrates an example showing how this algorithm works. In this example, for each object, two OSSs are randomly selected as potential targets and the one with the lighter load will be chosen according to the Algorithm~\ref{2}. 
In this specific example, we show that all objects are scheduled to the lighter load server based on their two random choices except for ob 3 that is scheduled to its default$\_$oss because its default$\_$oss (i.e., $OSS_3$) has a similar load compared to $OSS_5$ (the algorithm considers it is not worth of scheduling it to another location).
%
\subsubsection{ $n$-Level Two Random  ($n$LTR) algorithm }
In this scheduling algorithm, we first sort all storage servers in a descending order based on their probabilities in the statistic server log. Similar to TRH, this algorithm also considers randomized choices based on the server statistic log to mitigate the impact of stragglers. But, different from TRH, the randomized choices are no longer chosen from the top half of the sorted list of object storage servers; instead, a $n$-Level strategy is used. 
The $n$-Level strategy first divides the sorted storage servers list to the $K$ separate sections (where $K = 2^n$). At the same time, it also orders all pending I/O requests based on their lengths to form a descending order and divides them into the same number $K$ sections. To schedule, it will assign I/O requests in section $i$ into a randomly selected server in the $i$th section of the storage servers.
The way to divide storage servers and I/O requests is described as follows. Note that there is a core difference: we use the middle to divide storage servers to guarantee each section has the same number of storage servers; but we use the average to divide I/O requests to better utilize the size factor. 
Let's consider $S$ as the number of separate sections that we have so far, therefore:
		\begin{itemize}
			\item If $S =K$, no further action is required.
			\item If $S < K$,  the list of storage servers and I/O requests are divided to $2^l$  sections as follows, where $l$ represents level number :
			\begin{itemize}
			  	\item 	Middle/average element of each available section is chosen.
				  	\item   Each section is divided in two sections by using the middle/average element.
			  \end{itemize}
	 \end{itemize}
\begin{figure}[H]
	\begin{center}
		{\includegraphics[scale= 0.8,width=0.9\textwidth]{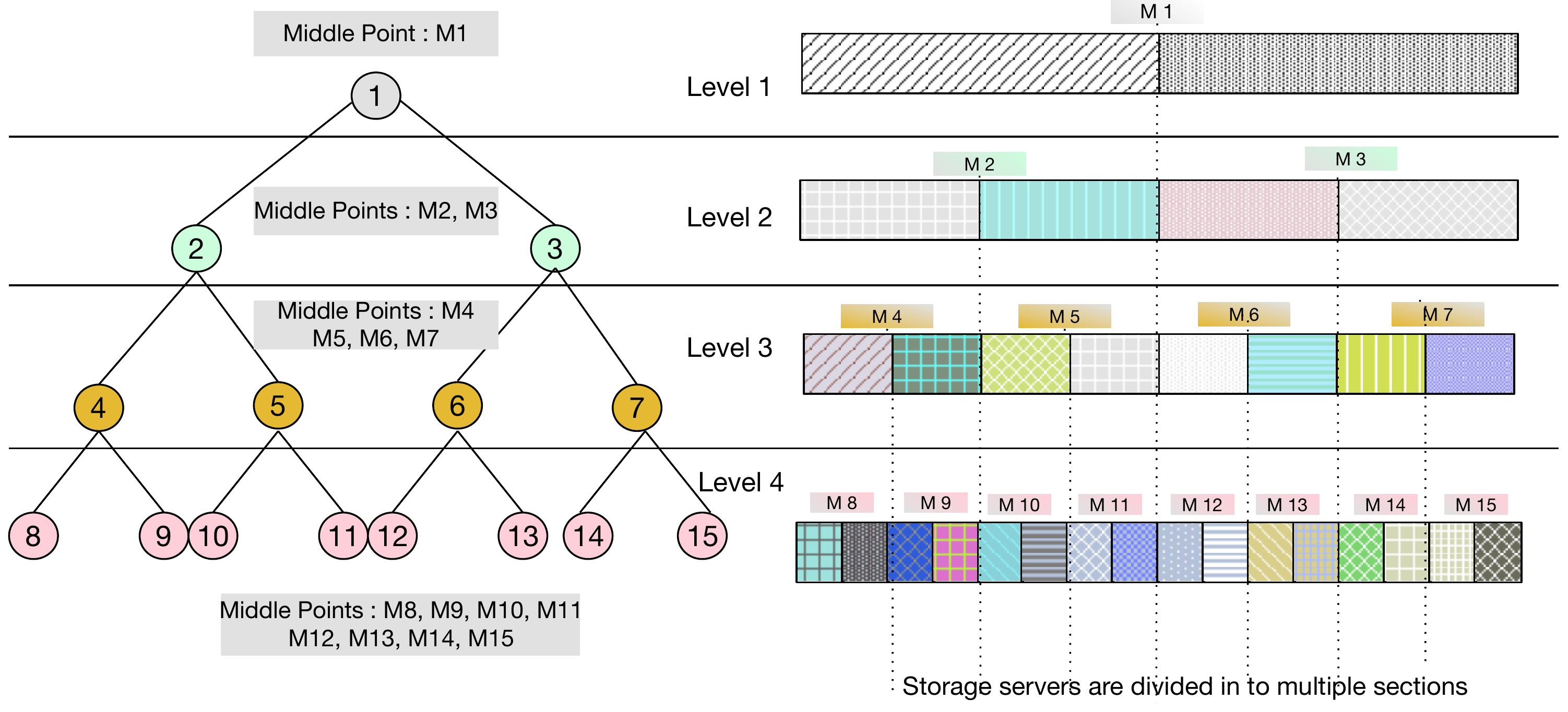}}
		\caption{Levels for dividing the storage servers.}
		\label{steps}
	\end{center}
\end{figure}	
We show an example of dividing storage server list in Figure~\ref{steps}. Here the chosen number of sections is $K$=16, and the number of levels is $n=\floor*{\log_2 K}  =  \floor*{\log_2 16} = 4$. All storage servers are divided in 16 sections.
As the figure shows, at the Level 1, the middle element of the list (i.e, M1) is used to divide the storage servers list in two sections.  At the Level 2, the middle elements of two sections are chosen (i.e, M2, M3) for dividing their corresponding sections. Therefore, at each level, for every section the middle element is chosen to divide it in two sections. This will continue until $K = 16$ sections have been created. 
The rationale of using $K$ sections in this algorithm is that it helps to spread the requests to be more balanced. 
Specifically, for a given I/O request, if its length is very small, it is more efficient to choose a storage server that might not be the lightest one but still has enough capability to process the I/O request. 
The detailed scheduling algorithm based on such sorted list is described in Algorithm~\ref{5}. We use the same threshold as in previous algorithms to decide whether choosing the calculated target$\_$oss or the default$\_$oss. 
\begin{algorithm}
	\caption{$n$-level Two Random ($n$LTR) Algorithm}\label{5}
	\begin{multicols}{2}
		\begin{algorithmic}[1]
			\Procedure{$n$LTR{(servers,requests,threshold)}}{}
			\State  sortedServers   = sort(servers);
			\State  requests  = sort(requests);
			\State index=0, i =0;  
			\State $level =1$;  
			\State K = NumberofSections; 
			\State  $n=\floor*{\log_2 K}$  \Comment{$n$=Number of levels;} 
			\While{(need\_schedule()}
			\State  default\_oss = requests[index] $\mod$ M;
			\While{$(level \ne n)$}
			\For{existSection()}
			\State M = findMiddleElement();
			\State MiddlePoints[i]=M;
			\State divideSection();
			\State i++;
			\EndFor	
			\State$level ++$;
			\EndWhile
			\State  T = compare(IOLength , MiddlePoints[]);	
			\State sectionNum = FindSection(T);
			\State  ross\_1=randomOssFromSectionNum(sectionNum);
			\State  ross\_2=randomOssFromSectionNum(sectionNum);
			\State target\_oss =MinLoad(ross\_1, ross\_2);
			\State benefit = load(default\_oss)- load(target\_oss);
			\If{benefit $\leq$ threshold}
			\State{$\Return$ target\_oss;}
			\Else
			\State{$\Return$ default\_oss;}
			\EndIf
			\State index++;
			\EndWhile
			\EndProcedure
		\end{algorithmic}
	\end{multicols}
\end{algorithm}
The $n$LTR algorithm allows users to decide an appropriate value for  $n$. In this research, we evaluate two initial cases $n=1$ and $n=2$ . Large $n$ clearly indicates higher computation complexity. But our evaluations did not show clear performance benefit of increasing $n$. For example, even the evaluation results for $n=3$ did not lead to significant performance benefit compared to $n=2$. However, the computation overhead to generate the sections for $n=3$ is greater than that of the $n=2$.
In general, the average computation overhead in $n$LTR algorithm is equal to the overhead required to generate sections for I/O requests and storage servers. Since generating sections ends up in a complete binary search tree, this means that  the $n$LTR algorithm can have the same big-O time/space complexity as the complete binary search tree. Therefore, the average computation overhead in $n$LTR algorithm is calculated as follows:
\begin{itemize}
	\item Space complexity: The space occupied to maintain middle and average points which is equal to 
	$ O(k) + O(k) = 2 O(k)$  for both average and worst case, where $k = 2^n$ 
	\item Time complexity: Is the time required to find (i.e., search) an appropriate section for a given I/O request which is equal to 
	  $O(k\log_2 k)$  for average case, and $O(n)$  in worst case.
\end{itemize}
Hence, we suggest these two values (i.e., $n=1$ and $n=2$) should be used in most cases.

\section{Evaluation and Analysis}
\label{s4}
\begin{figure}[!tbp]
	\centering
	\begin{minipage}[b]{0.48\textwidth}
		{\includegraphics[width=\linewidth]{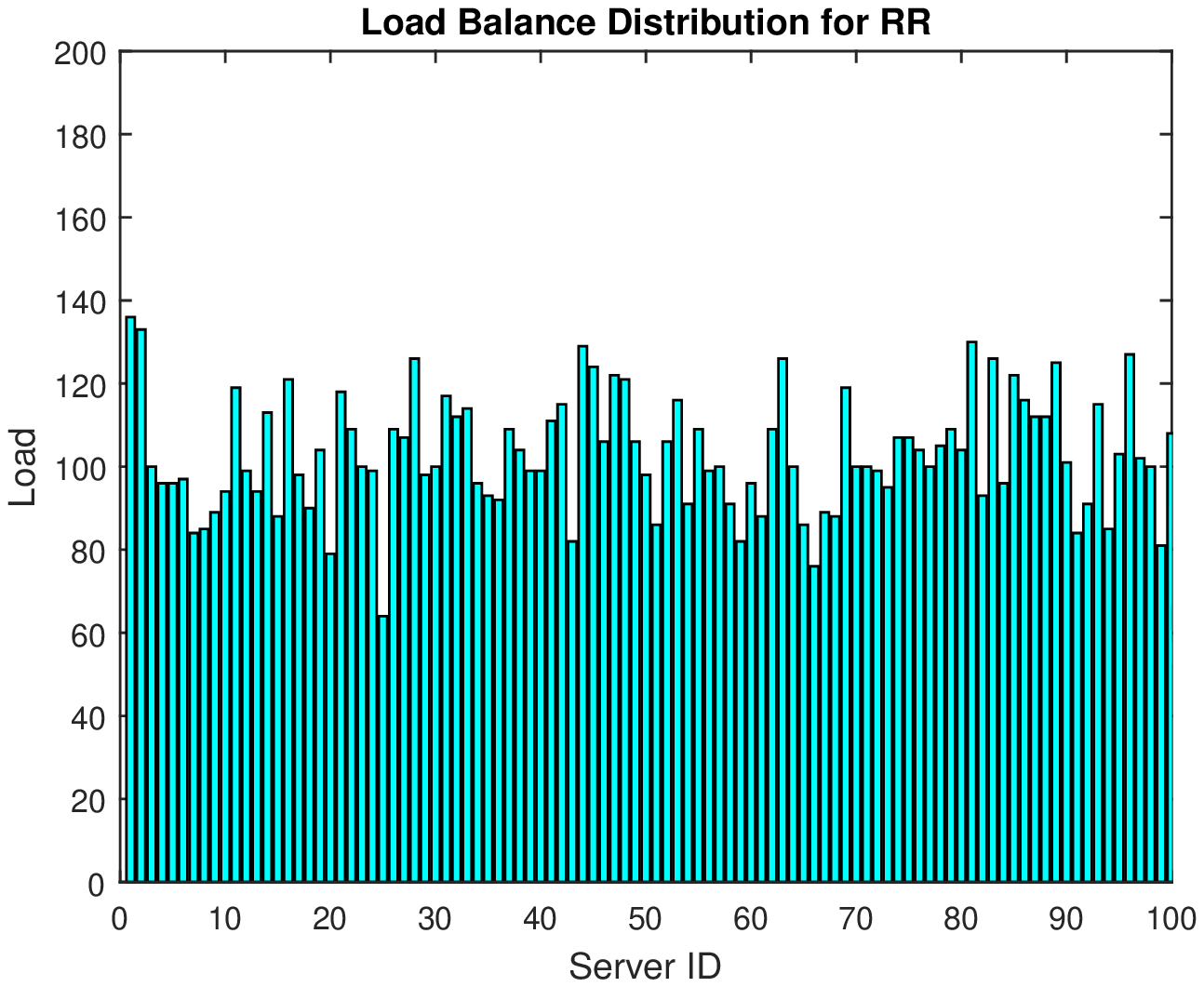}}
		\caption{Load distribution with the RR scheduling algorithm.}
		\label{r1}
	\end{minipage}
	\hfill
	\begin{minipage}[b]{0.48\textwidth}
		{\includegraphics[width=\linewidth]{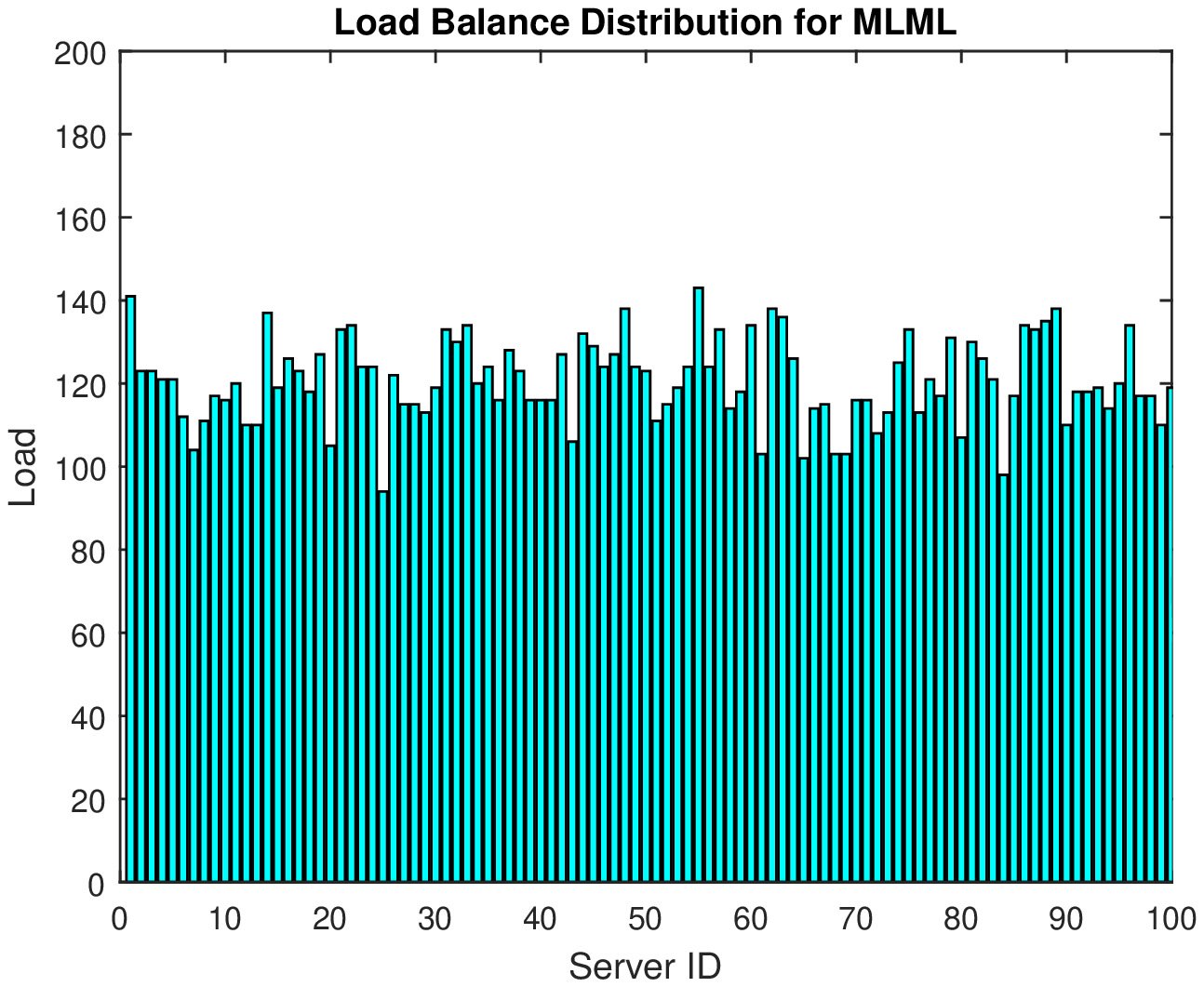}}
		\caption{Load distribution with the MLML scheduling algorithm.}
		\label{r2}
	\end{minipage}
\end{figure}

\begin{figure}[H]
	\begin{minipage}[b]{0.45\textwidth}
		{\includegraphics[width=\linewidth]{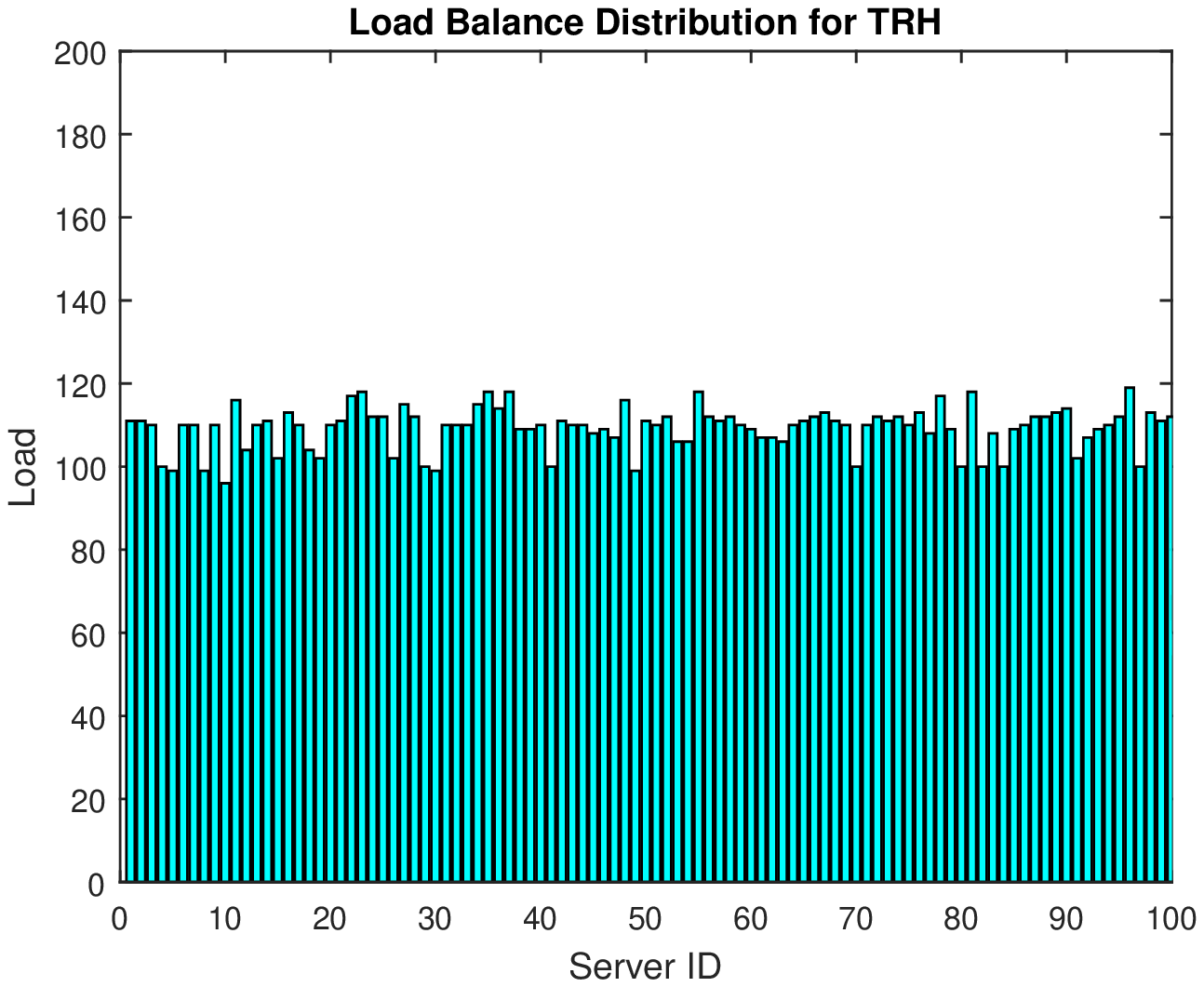}}
		\caption{Load distribution with the TRH scheduling algorithm. (First run)}
		\label{r3}
	\end{minipage}
	\hfill
	\begin{minipage}[b]{0.42\textwidth}
		{\includegraphics[width=\linewidth]{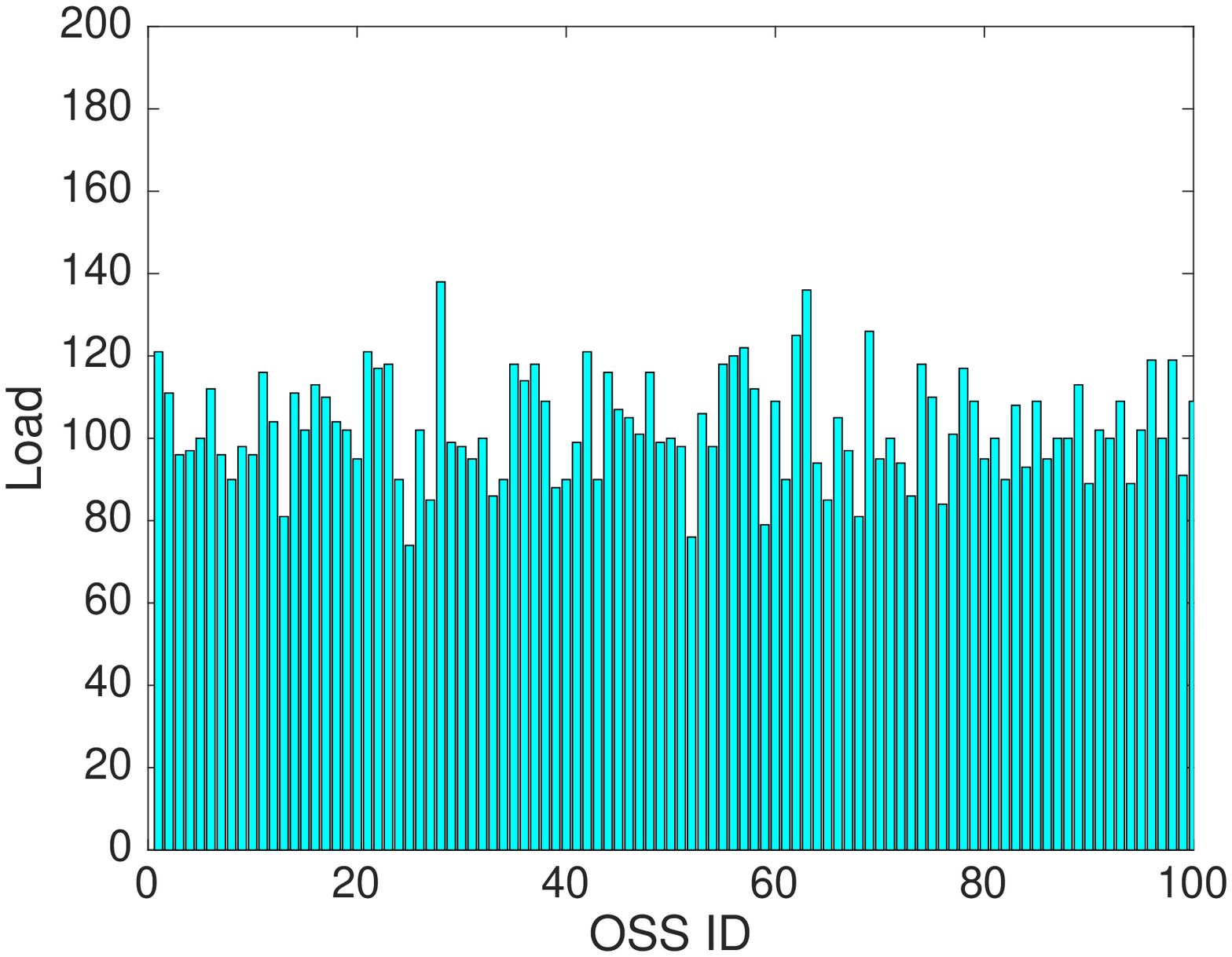}}
		\caption{Load distribution with the TRH scheduling algorithm. (Second run)}\cite{tavakoli2016log}.
		\label{load3}
	\end{minipage}
	
\end{figure}

\begin{figure}[H]
	\begin{minipage}[b]{0.45\textwidth}
		{\includegraphics[width=\linewidth]{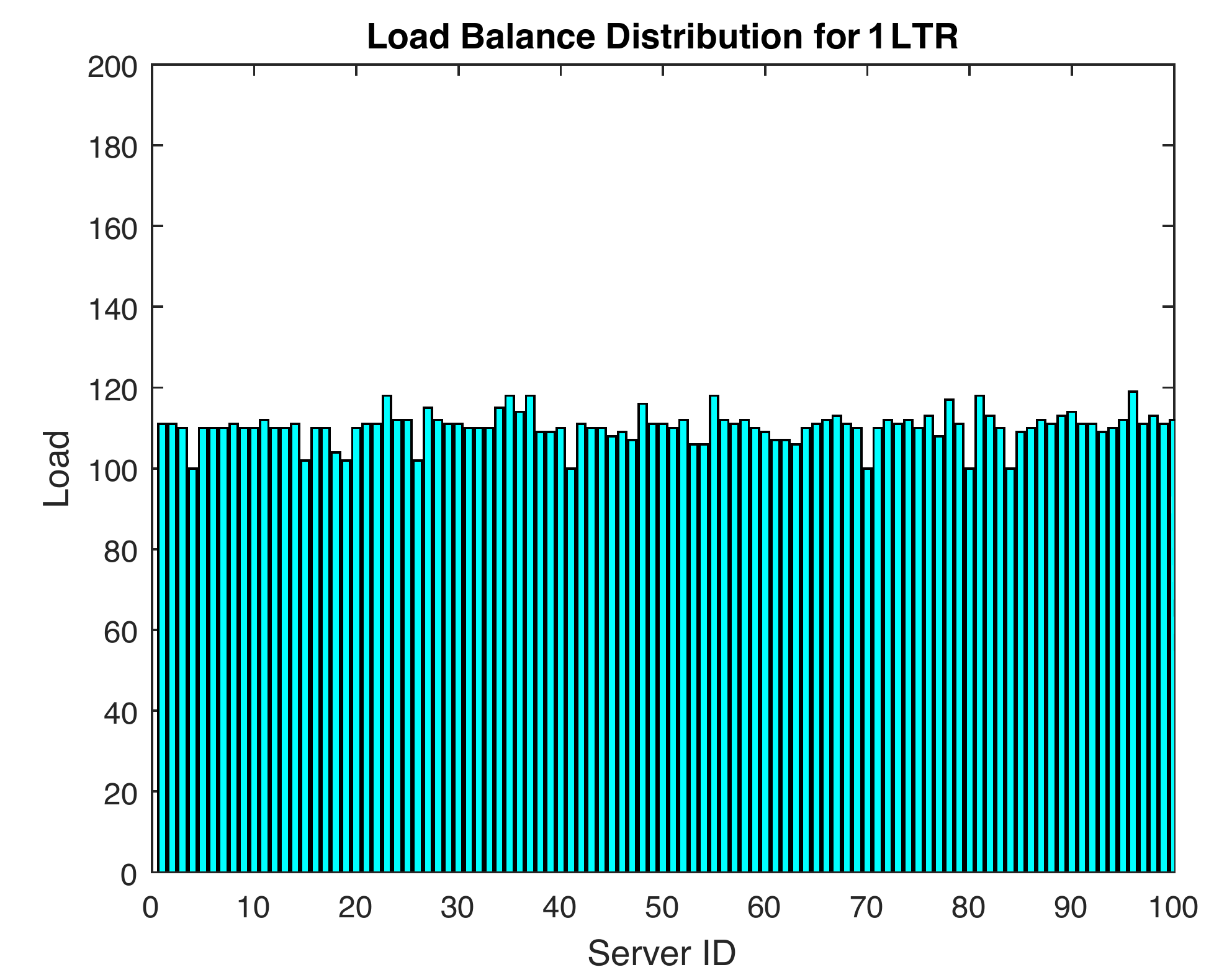}}
		\caption{Load distribution with the 1LTR scheduling algorithm.}
		\label{r4}
	\end{minipage}
	\hfill
	\begin{minipage}[b]{0.48\textwidth}
		{\includegraphics[width=\linewidth]{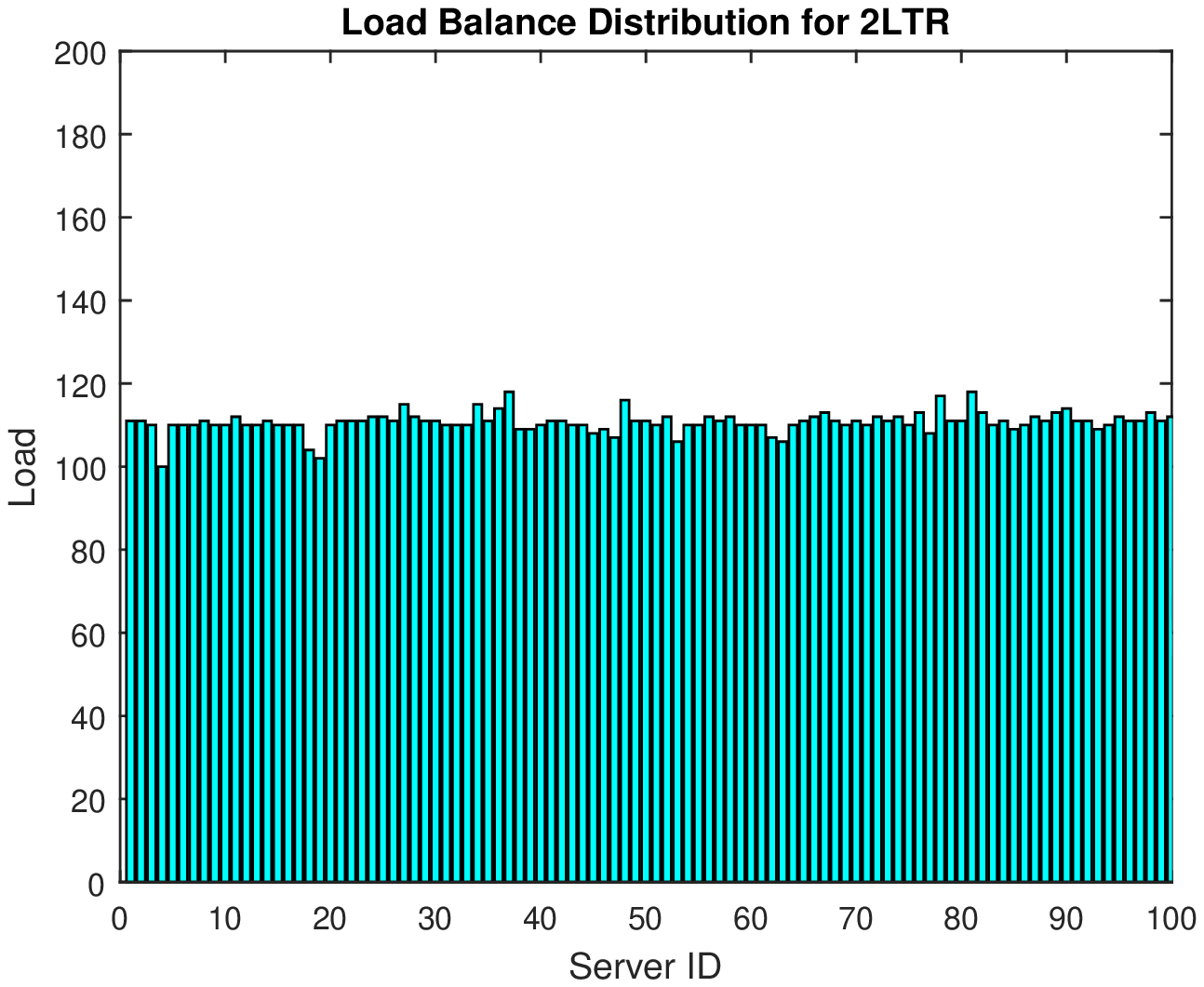}}
		\caption{Load distribution with the 2LTR scheduling algorithm.}
		\label{r5}
	\end{minipage}
	
\end{figure}

\begin{figure}[H]
	\begin{center}
		{\includegraphics[width=0.9\linewidth]{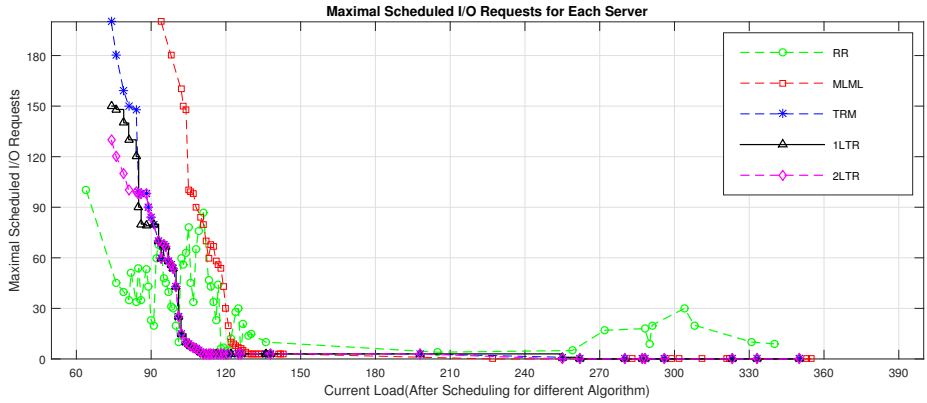}
			\caption{Comparison of RR, MLML, TRH, 1LTR, and 2LTR scheduling algorithms.}
			\label{r6}}
	\end{center}
\end{figure}

We have evaluated the proposed I/O scheduler and scheduling algorithms using simulation. 
In the simulation, we used synthetic workloads generated based on real-world traces. We constructed the synthetic workload based on combining three different types of I/O requests: large I/O (each request is greater than $O(10MB)$), medium I/O (each request is between 4MB and 10MB) and the small one (less than 4MB). These three categories of requests are chosen based on our observation of the typical scientific applications running on HPC systems. 
In addition, the initial I/O loads of all OSSs are generated based on a normal distribution with a small standard deviation, which simulates a roughly even workload distribution at the beginning.
The simulated environment is an HPC cluster with 100 object storage servers and 200 compute nodes. In all test cases, we run the simulation 100 times  and calculate the average load of each OSS with the proposed log-assisted straggler-aware scheduler and with the MLML, TRH, 1LTR, and 2LTR scheduling algorithms, respectively. The total number of I/O requests simulated was 2,000, with various sizes for each request; for instance, the total data written for all medium I/O case can be up to 20GB. The total data written for all large I/O case can be between $O(20GB)$ and $O(2TB)$ depending on the size of each request.
We consider the round-robin (RR) algorithm as the base-line case and compare it with the proposed straggler-aware scheduler with three algorithms. In Fig.~\ref{r1}, we first illustrate the load of each server after scheduling all I/O requests using the round-robin policy. The $x$-axis shows the storage
servers, and the $y$-axis represents the load of each storage server (i.e, write data size in MB).
Fig.~\ref{r2} plots the load distribution of storage servers with the log-assisted straggler-aware scheduler after scheduling all requests using the MLML algorithm. As the figure shows, since the MLML algorithm rotates the requests to servers based on their loads, this algorithm is able to achieve a better load balance across different storage servers compared with Fig.~\ref{r1}.
We further show the result of load distribution with the TRH algorithm in Fig.~\ref{r3}, and in Fig.~\ref{load3} \cite{tavakoli2016log}. Since this algorithm randomly chooses two servers from the top half of the lighter loaded storage servers and selects the better one, it can effectively avoid stragglers to achieve better balance compared with the round-robin strategy. However, due to the randomness behavior it may have larger/smaller variances than that of the MLML algorithm. Fig.~\ref{r3} illustrates the case where TRH performs better than MLML but Fig.~\ref{load3} represents the case that TRH achieves worse balance compared with the MLML.
We further show the result of load distribution with the 1LTR and 2LTR  algorithm in Fig.~\ref{r4} and Fig.~\ref{r5}. Since these algorithms randomly choose two servers from multiple sections, they can effectively avoid stragglers to achieve better balance compared with the round-robin strategy. 
To further validate and verify the effectiveness of the proposed log-assisted straggler-aware I/O scheduler, in this series of evaluations, we manually inject a number of storage servers as stragglers. We expect the straggler-aware scheduler to be able to avoid hitting existing stragglers and also avoid generating new stragglers. 
In these tests, we assigned 10\% of the total object storage servers as stragglers by adding extra loads on those selected servers. Specifically, we injected 5 times more load compared with the average loads assigned on other storage servers.
We report the result in Fig.~\ref{r6}. In this figure, the $x$-axis indicates all the possible loads of storage servers after scheduling; the $y$-axis shows the maximal IO requests scheduled onto storage servers that have the corresponding load as the $x$-axis shows. 
We only show the maximal I/O requests since there might be multiple servers having the same load after scheduling, and the one receiving the most I/O requests actually determine the overall performance as we have described. 
An effective straggler-aware scheduler should be capable of avoiding stragglers; in other words, the number of I/O requests falling into the slow storage servers (stragglers) should be close to zero. Based on the simulation results reported in Fig.~\ref{r6}, the log-assisted straggler-aware scheduler with MLML, TRH, 1LTR, and 2LTR algorithms are capable of achieving that since they do not schedule any I/O request onto servers with load greater than 200 in our test cases. However, the RR strategy still schedule I/O requests to highly loaded servers (over 300). 
Between the MLML and TRH algorithm, we can notice that the TRH scheduling algorithm tends to place loads on lighter loaded servers (the curve is on the left side of the MLML curve). This behavior is due to the fact that randomized scheduling choices introduce more options for selection and better promote to spread the requests.
Also, between 1LTR and 2LTR, we can notice that they are largely overlapped indicating they provide the similar capability of achieving the balanced I/O. Together with the results plotted in Fig~\ref{r4} and Fig~\ref{r5}, we show that, selecting $n =2$ is good enough for mitigating stragglers and providing a load balanced storage server. 
These simulation results show that using the client-side server statistic log and the proposed scheduling algorithms can avoid storage server stragglers and improve the scheduled workload balance among storage servers.
%
\section{Related Work and Comparison}
\label{s5}
I/O scheduling has been extensively studied in earlier research efforts. The straggler problem has been well acknowledged as one of critical issues in I/O scheduling. Current research is heavily focused on trying to avoid stragglers, including our previous work of an I/O scheduler based on a two-choice randomized scheduling that dynamically places write operations to mitigate the straggler problem in high-end/high-performance computing systems~\cite{dai2014two}. This paper is different from two-choice approaches as we avoid the probing latency through using client-side logs.
ADIOS~\cite{lofstead2010managing} introduced an adaptive I/O method to reduce the I/O variability within a single application. It gathers all processes inside the application in a group and each group writes in a particular storage server.
CALCioM~\cite{dorier2014calciom} mitigates I/O interferences in HPC systems through a cross-application coordination. It allows different applications to communicate and coordinate their I/O strategies to avoid congestion. Our study is different from it as we introduce a client-side scheduler that can transparently identify those patterns and coordinate to mitigate stragglers and hence alleviate the interferences.
There are also a series of research focusing on identifying I/O access patterns of applications and using such information to mitigate stragglers and improve the performance~\cite{Luyry14,dydr14,he2013acceleration}. Our research is different from them as we do not need to know the details about each application; instead, by checking the requests and previous served I/O logs, we are able to dynamically schedule the I/O requests to mitigate stragglers.
The two-choice algorithm is also used in Sparrow, a distributed task scheduler~\cite{ousterhout2013sparrow}. It uses multiple distributed schedulers on a cloud-based platform to schedule a large number of short tasks. Sparrow avoids the straggler problem by probing multiple nodes for each task and scheduling that task on the node with the least overhead. The Sparrow scheduler is specifically designed for cloud systems and workloads though. 
Another straggler mitigation technique is introduced in Dolly with a proactive straggler mitigation approach~\cite{ananthanarayanan2012let}. It uses a proactive database provisioning scheme and leverages virtual machine cloning in cloud platforms. It accomplishes this by using the power-law distribution of job sizes to launch different clones of each task. Furthermore, it attempts to predict stragglers by waiting and observing the system behavior before scheduling a task. 

Yet another straggler mitigation technique is presented in Mantri~\cite{ananthanarayanan2010reining}, which moves tasks from straggling resources and assigns them to less overloaded resources. It monitors tasks and attempts to avoid stragglers based on their causes. It accomplishes this goal in three primary steps. First, speculative copies of straggler tasks are executed while being aware of resource constraints and work imbalance. Second, all tasks are placed based on the locations of their data sources. Last, once a task is completed, its output (intermediate data) is replicated on other resources with lighter workloads.

One of the widely used techniques for mitigating the impact of stragglers is a speculative execution technique. Speculative execution means that, for those tasks that are stragglers or are likely to be stragglers, extra copies of tasks are executed. Among these copies, the one that finishes first is chosen~\cite{ananthanarayanan2014grass,dean2008mapreduce, ananthanarayanan2010reining}.
Many systems and studies use speculative executions to address the straggler problem~\cite{ ananthanarayanan2013effective,zaharia2008improving,ananthanarayanan2010reining, ananthanarayanan2014grass}, including a cloning approach~\cite{xu2015task,ananthanarayanan2013effective} and a straggler-detection-based approach ~\cite{ chen2014improving,isard2007dryad}. The primary difference between these two approaches is how/when they launch their speculative task copies. In the cloning approach, the initial task is scheduled in parallel with extra copies, and the task that finishes first is used for the next scheduled task. In the straggler-detection based approach, the extra copies are only launched if stragglers are detected. 
One speculative task scheduler, LATE~\cite{zaharia2008improving}, implicitly assumes that cluster nodes are homogeneous and linearly make progress. It uses these assumptions to speculatively re-execute tasks that appear to be stragglers. In fact, it focuses on reducing the response time of scheduling by speculating which running task can be overtaken. Any task that may be overtaken will then be re-executed to avoid stragglers.
Hopper~\cite{ren2015speculation}, is another example of a job scheduler that uses speculative execution to mitigate the straggler problem.
The key idea of Hopper is that it must predict the speculation requirements of jobs and then dynamically launch extra copies on alternative machines.
The downside of speculative execution is that, while it mitigates the impact of stragglers, it consumes valuable system resources due to redundant executions. Furthermore, it does not coordinate resources for concurrent jobs well.

Recently, an interesting work is introduced  in~\cite{xie2017output} in which authors show that a small proportion of storage servers ($< 20\%$) are straggler at any	given interval, but	that stragglers	are	transient which means any storage server that is currently a straggler, it will return to normal operation within 2 minutes. Any non-straggler storage will become straggler within 10 minutes. Furthermore, in~\cite{xie2017predicting} authors show that  while application is running, all I/O will trend to a long-term average performance for that sized I/O.

In this research, we introduce a log-assisted straggler-aware I/O scheduling. It is different from these existing studies as it addresses the storage server straggler problems in object-based storage systems. It is specifically designed for HPC systems where high concurrency is the norm and uses client-side logs to reduce probing messages that are required in two-choice or similar scheduling algorithms. It avoids redundant executions as required in the class of speculative techniques. To the best of our knowledge, this study is a first research effort that introduces a log-assisted straggler-aware scheduling approach.
\section{Conclusion and Future Work}
\label{s6}
Many high-end/high-performance computing applications have become highly data intensive and the needs of highly efficient storage systems to better support scientific discovery substantially grow over years. In this research, motivated by the imperative needs of better dealing with storage server stragglers (servers that take much longer time in responding to I/O requests than other servers due to transient overloaded requests, imbalanced accesses, or even hardware/software transient failures), we introduce a new log-assisted straggler-aware I/O scheduler.
We have presented the idea, design, and evaluation of such straggler-aware I/O scheduling strategy. 
We have introduced a client-side server statistic log to maintain I/O requests and servers’ status so that the I/O scheduler client can make an optimized scheduling decision when stragglers exist, without incurring expensive probing messages as in the existing straggler-aware I/O scheduling methods. We presented three new straggler-aware scheduling algorithms based on the proposed client-side statistic logs: Max Length-Min Load (MLML) algorithm, Two Random from Top Half (TRH) algorithm, and ($n$-Level) Two Random algorithm. Our evaluations confirmed that these three scheduling algorithms can reduce the impact of existing stragglers and avoid generating new stragglers, which shows the benefit of log-assisted straggler-aware scheduling scheme for object-based storage systems.  

We plan to investigate and integrate the proposed I/O scheduling scheme into existing real parallel file systems. In addition, while the current simulation shows the solution improves the I/O performance, we believe that more optimization can be made with the statistic logs, and we plan to further investigate the possibilities in the future.  
\section*{Acknowledgement}
This material is based upon work supported by the National Science Foundation under grant CCF-1409946 and CNS-1338078. 
%
\section*{References}
\bibliographystyle{IEEEtran}
\bibliography{bib}
\end{document}